# Tuning to non-veridical features in attention and perceptual decision-making: An EEG study

**Stefanie I. Becker, Zachary Hamblin-Frohman, Hongfeng Xia & Zeguo Qiu**

School of Psychology, The University of Queensland

Brief title: Tuning to non-veridical features

Word count: 8,358

Corresponding author information:

Dr Stefanie Becker

School of Psychology

The University of Queensland

Brisbane

Australia

Email: s.becker@psy.uq.edu.au

Ph: +61 449 883870

## Abstract

When searching for a lost item, we tune attention to the known properties of the object. Previously, it was believed that attention is tuned to the veridical attributes of the search target (e.g., orange), or an attribute that is slightly shifted away from irrelevant features towards a value that can more optimally distinguish the target from the distractors (e.g., red-orange; *optimal tuning*). However, recent studies showed that attention is often tuned to the relative feature of the search target (e.g., redder), so that all items that match the target's relatively features equally attract attention (e.g., all redder items; *relational account*). Optimal tuning was shown to occur only at a later stage of identifying the target. However, the evidence for this division mainly relied on eye tracking studies that assessed the first eye movements. The present study tested whether this division can also be observed when the task is completed with covert attention and without moving the eyes. We used the N2pc in the EEG of participants to assess covert attention, and found comparable results: Attention was initially tuned to the relative colour of the target, as shown by a significantly larger N2pc to relatively matching distractors than a target-coloured distractor. However, in the response accuracies, a slightly shifted 'optimal' distractor interfered most strongly with target identification. These results confirm that early (covert) attention is tuned to the relative properties of an item, in line with the relational account, while later decision-making processes may be biased to optimal features.

**Tuning to non-veridical features in attention and perceptual decision-making**

In everyday life, we spend a lot of time looking for things, such as a lost phone, wallet, set of keys, or trying to spot a friend in a crowded restaurant or find our car in a parking lot. Knowing the visual properties of a sought-after item can immensely help our search, which is testament to our ability to 'tune' visual attention to particular features in a top-down, goal-driven manner (Wolfe, 1994; see also Hout & Goldinger, 2015; Soto et al., 2008). Theories describing how we can use our knowledge to tune attention can be broadly classified into 'representational theories' vs. 'analytical/computational theories' (c.f., Becker, Martin & Hamblin-Frohman, 2019). Representational theories propose that we have a mental representation of the target, a *target template*, which can guide attention to items in the visual field that match the target template (e.g., Duncan & Humphreys, 1989; Hout & Goldinger, 2015; Soto et al., 2008). The target template will usually be derived from memory, and can reside in visual short-term memory (VSTM) or long-term memory (LTM; e.g., Carlisle, Arita, Pardo & Woodman, 2011; Olivers et al., 2011; Woodman et al., 2007). When searching for an item, memories residing in LTM will usually be uploaded to VSTM, and this process may be necessary for the target representation to guide visual attention (e.g., von Morselaar, Theeuwes & Olivers, 2014; Olivers, Peters, Houtkamp & Roelfsma, 2011).

The analytical / computational theories usually describe guidance in terms of our ability to up- and/or down-modulate the response gain of sensory neurons that can signal the location of particular features (e.g., colour) in the visual field (i.e., *feature maps*; e.g., Wolfe, 1994). Tuning attention to red in order to, for example, find a red car in a parking lot, would result in up-modulating sensory neurons that respond to red, so that these neurons dominate the collective neuronal response and guide attention first to all the red items in the visual field (e.g., Wolfe, 1994, 1998, 2022; see Treisman & Sato, 1990 for an inhibition account).

Representational and analytical theories are not incompatible. Even if '*tuning attention*' may, strictly speaking, most closely mean 'activating a target template in VSTM' in representational theories versus 'modulating the response gain of sensory neurons' in analytical theories, the two accounts are not mutually exclusive. In fact, analytical theories may complement representational views, by providing a plausible mechanism for how representational contents can guide visual attention (e.g., Eimer, 2014; but see also Desimone & Duncan, 1995).

Importantly, both representational and analytical theories typically assume that we tune attention to the veridical features of a sought-after item. As an object typically has multiple features (e.g., colour, shape, size, etc.), some accounts may predict that we would tune attention to the most useful or 'diagnostic' feature of the target (e.g., Boettcher, van Ede & Nobre, 2020; Wolfe, 1994, 2021). Still, the feature(s) would closely correspond to attributes the target actually has (i.e., veridical features). There are only two theories that predict tuning to (different) non-veridical features in particular conditions: the *optimal tuning account* and the *relational account*.

According to the optimal tuning account, we would indeed usually tune attention to the exact target feature value (in line with the mainstream views), except when the target is surrounded by nontarget items that are very similar to the target (e.g., Navalpakkam & Itti, 2007). In this condition, tuning attention to the target would result in a poor signal-to-noise ratio (SNR), as it would up-modulate the response gains of both the target and the nontargets. Hence, to increase the SNR, attention will be tuned to a feature that is slightly shifted away from the nontargets to a slightly exaggerated target feature that is more optimal for distinguishing the target from the nontargets. For example, if the target is orange and surrounded by yellow-orange nontargets, attention would be tuned to a slightly more reddish orange, to maximise the difference between the neural response to the target vs. the nontargets (e.g., Navalpakkam & Itti, 2007; Yu & Geng, 2019). In line with the predictions of the optimal tuning account, it has been found that observers will often report a distractor with an exaggerated target feature as the target, if (and only if) the target was presented among very similar nontargets on the majority of trials (e.g., Navalpakkam & Itti, 2007; Scolari & Serences, 2009, 2010; Yu & Geng, 2019).

A different view has been proposed by the *relational account*. According to this view, attention is often not tuned to a particular feature value like a specific colour at all. Instead, the visual system assesses the dominant features in the visual scene and tunes attention to the relative feature that best discriminates the target from (most of) the surrounding items. For instance, when an orange target is presented among mostly yellow items, attention will be tuned to all redder items or the reddest item. As a consequence, the reddest item will be selected first, followed by the next-reddest item, and so forth (Becker, 2010).

Deviating from the optimal tuning account, tuning to relative features is predicted to occur independently of nontarget similarity, and can lead to selection of vastly different colours rather than being limited to a range of similar feature or items with a slightly different (shifted) feature value (e.g., York & Becker, 2020). According to the relational account, attention will only be tuned to a particular feature value when the target cannot be found by tuning to the relative feature, as for example when an orange target is surrounded by equal numbers of red and yellow items (e.g., Becker, Harris, Venini & Retell, 2014; Harris, Remington & Becker, 2013; Schoenhammer, Becker & Kerzel, 2020).

Two studies subsequently tested whether attention is tuned optimally or relationally (Hamblin-Frohman & Becker, 2021; Yu, Hanks & Geng, 2022). Both used the first eye movement in visual search trials to probe into processes that guide early visual attention, as the first eye movement is usually executed quite early (i.e., within 150-250ms after display onset) and is commonly regarded as a suitable marker for processes that guide visual attention (e.g., Ramgir & Lamy, 2021; Zhaoping & Frith, 2011). Optimal vs. relational tuning was tested in displays containing a target with a constant, known colour (e.g., orange) which was presented among either similar (e.g., yellow-orange) or dissimilar (e.g., yellow) nontargets (in different blocks), plus an irrelevant distractor that could have a range of different colours (e.g., ranging from full red to yellow). Critically, the distractor colours included 3-4 relatively matching colours that systematically differed in similarity to the target. The results showed that observers were equally likely to select all relatively matching distractors with the first eye movement, regardless of whether the distractor was similar or dissimilar to the target colour, in line with the relational account.

Evidence for optimal tuning was only found in a late measure, viz., the accuracies in responding to the target: When the target was similar to the nontargets, observers were more likely to report a distractor that had a slightly exaggerated target colour (i.e., was similar to the target and slighly shifted away from the nontarget colour; e.g., red-orange). This effect was not observed when the target was dissimilar from the nontargets, as predicted by the optimal tuning framework.

Given that the optimal tuning effects were observed only rather late in the visual search trials, it was concluded that optimal tuning does not guide attention, but appears to guide perceptual decision-making *after* an item has been selected (when the task requires very fine-grained perceptual decision-making; e.g., Scolari & Serences, 2009, 2010). Attention, by contrast, is guided by a much broader, relational target 'template' that can include a range of different colours that may even cross the colour boundaries (Yu et al., 2022; see also York & Becker, 2020).

While these results mark important progress in understanding the factors and mechanisms guiding attention vs. perceptual decision-making, it should be noted that both studies used eye movements to index attentional guidance. Eye movements are regarded as a very good marker for early attentional selection, not only because they are completed quite rapidly, but also because an eye movement to a location is usually preceded by a covert attention shift to the location (e.g., Deubel & Schneider, 1996). Still, we can ask whether the results generalise to tasks that require no eye movements, and allow only covert attention shifts. It is well-known that we can shift covert attention without moving our eyes, and covert attention may be deployed differently when it occurs without an eye movement vs. when it precedes a pre-planned eye movement (e.g., Wu & Remington, 2003).

Moreover, in previous eye movement studies, the search stimuli were typically quite widely spaced out, to encourage eye movements and optimise the conditions for measuring them accurately (e.g., Hamblin-Frohman & Becker, 2021). It is possible that these displays encourage adopting a broader, relational target template, whereas this may not be true for displays with more densely packed stimuli that are closer to fixation (as is typical for most studies assessing only accuracies and RTs). Thus, it is still an open question whether testing the relational account against optimal tuning would show the same results when selection is based on covert attention, in more densely populated displays and with search stimuli that are closer to the fovea.

To address this question, the present study used the N2pc in the electroencephalogram (EEG) of participants to track covert attention shifts. The N2pc has been established as a marker for covert attention (e.g., Eimer, 1998; Luck et al., 1997; Woodman et al., 2009), and is reflected in an increased negativity contralateral to the attended side in posterior electrodes that occurs in a time window of 200 – 300ms after stimulus onset (e.g., Eimer, 1998).

Of note, the N2pc may not index attentional guidance proper or transient shifts of covert attention. It has been suggested that instead, the N2pc may indicate slightly later processes of attentional engagement or attentional selection that commence shortly after covert attention has shifted to an item (e.g., Ramgir & Lamy, 2022; see also Kiss et al., 2008; Mazza & Carramazza, 2011; Zivony et al., 2018). This may also explain why salient, irrelevant stimuli typically fail to generate a significant N2pc (Jannati, Gaspar & McDonald, 2013; McDonald, Green, Jannati, & Di Lollo, 2013; but see Burra & Kerzel, 2013): As these stimuli are typically very dissimilar from the target, they can be rejected quite quickly, without much attentional engagement or feature analysis (e.g., Zivony & Lamy, 2018).

For the purpose of the present study, the possibility that the N2pc may indicate slightly later processes should not present a difficulty, as both relational and optimal tuning are supposedly goal-driven and thus, should have longer-lasting effects and lead to attentional engagement rather than just having a fleeting, short-lived effect on attention (e.g., Becker, 2010; Navalpakkam & Itti, 2007). Correspondingly, previous EEG studies have shown a significant N2pc in response to irrelevant items that matched the relative colour of the target (e.g., Martin & Becker, 2018; Schoenhammer, Grubert, Kerzel & Becker, 2016). As these studies did not distinguish between optimal vs. relatively matching distractors, it is unclear if covert attention was guided by optimal or relational tuning. However, as the stimuli generated a significant N2pc, the N2pc seems well-suited to investigate whether covert selection follows optimal tuning or relational tuning (Hamblin-Frohman & Becker, 2021; Yu et al., 2022).

Another component in the EEG that is relevant for the present study is the Pd. The Pd is reflected in an opposite, *positive* contralateral deflection in the EEG of participants and has been linked to suppression or inhibition of the distractor (e.g., Gaspelin & Luck, 2018; Hickey, DiLollo & McDonald, 2009). Uncharacteristically for ERPs, the Pd does not appear to have a fixed time window but can appear either in a similar time window as the N2pc (e.g., 200 – 300ms post stimulus onset) or later (e.g., 300 – 400 ms post stimulus; e.g., Papaioannou & Luck, 2019; Sawaki & Luck, 2010; Sawaki, Geng & Luck, 2012). These findings have been taken to show that an irrelevant distractor can be inhibited prior to being selected (i.e., preventing selection) or after selection, to facilitate re-orienting to the target (e.g., Sawaki et al., 2012; but see Kerzel & Burra, 2020; Livingstone, Christie, Wright & McDonald, 2017 and Schoenhammer, Becker & Kerzel, 2020). In the present study we analysed both the N2pc and possible Pd in response to the distractors, including in later time windows (after a possible N2pc).

The aim of the present study was to test whether a covert attention task would mimic previous eye movement results, viz., whether covert attention would be allocated to all relatively matching stimuli, following the relational account, while perceptual decision-making would follow optimal tuning. To that aim, the target, nontarget and distractor colours were chosen from equiluminant colours that systematically varied between green and blue (see Fig. 1A). The target was always greenish-blue (olive). For different participants, the target was either embedded among turquoise nontargets, so that the target was bluer than most of the search stimuli, or the olive target was presented among aqua nontargets, so that the target was *greener* than the nontargets (randomly determined; see Fig. 1A). The target and nontargets were triangles, and participants had to select the target covertly (i.e., without moving the eyes) and to report the direction of the target triangle with a button press (up/down).

To distinguish between relational and optimal tuning, we included an irrelevant square distractor that could have one out of six different colours. blue, aqua, olive, turquoise or green. The colours were relabelled for analysis and display purposes, depending on the target condition (i.e., nontarget colours; Fig 1A). In the greener target condition (olive target among aqua nontargets), attention should be tuned to all greener items, or the greenest item, according to the relational account and previous results (Hamblin-Frohman & Becker, 2021; Yu et al., 2022). Hence, green was labelled the *relational* colour. According to the optimal tuning account, attention should be tuned to a colour slightly shifted away from the target colour (as the nontargets were all similar to the target). Hence, turquoise was labelled the *optimal* colour. In the bluer target condition (olive target among turquoise nontargets), blue was labelled the relational colour and aqua was labelled the optimal colour. Olive was always labelled the *target* colour; aqua and turquoise were labelled *nontarget* colours (depending on the condition), and green and blue were labelled as *opposite* colours when they were not the relational colour, as they differed in the opposite direction from the target.

According to the optimal tuning account, the optimal distractor should attract attention most strongly, followed by the target-coloured distractor, whereas the remaining distractors should not attract attention (e.g., Navalpakkam & Itti, 2007; see Fig 1B). According to the relational account, all relatively matching distractors should attract attention; so that we would expect equally strong attentional capture for the relational and optimal distractors, slightly weaker capture by the target-coloured distractor and no capture by any of the remaining distractors.

Covert attention to the distractors was assessed by the mean N2pc amplitude to the distractor. If the present study replicates previous findings from eye tracking studies (Hamblin-Frohman & Becker, 2021; Yu et al., 2022), the results should follow the relational account, with larger N2pc amplitudes for the relational and optimal distractors than for the target-coloured distractor (and no N2pc for the remaining distractors; see Fig 1B). Evidence for optimal tuning should only be obtained in later perceptual decision-making processes.

To accurately measure perceptual decision-making processes, we interleaved probe trials with the visual search trials (as in previous studies; see Hamblin-Frohman & Becker, 2021; Navalpakkam & Itti, 2007; Yu et al., 2022). The probe displays contained two colours; the target colour plus one of the possible distractor colours, which were presented only briefly and backward-masked (see Fig. 2). The participants' task was to indicate the location of the target colour by pressing a button (left/right). If perceptual decision-making is tuned to a more optimal colour to help with fine-grained discriminations, then participants should be less accurate in reporting the target probe with the optimal colour than all other colours (Hamblin-Frohman & Becker, 2021; Navalpakkam & Itti, 2007).

We did not analyse the N2pc or Pd in response to the probe displays, as these were presented only on a small portion of trials, resulting in insufficient trial numbers. In addition, the probe displays always contained a target on one side of the display, which is not ideal for measuring lateralized ERPs such as the N2pc or Pd in response to the distractor colour, as the target will strongly compete for attention with the distractor (e.g., Kiss & Eimer, 2008).

To optimize the conditions for measuring the N2pc and Pd to the different distractors in visual search, we created displays in which the target was presented on the midline (and thus, would not evoke a lateralized potential) and the distractor was presented on the right or left side of the screen (see Fig. 1, bluer target). To prevent that the target location would become predictable, we interspersed these trials with trials in which the target was lateralised, and a distractor was either absent or presented on the midline (see Fig. 1, greener target). These trials were excluded from all EEG analyses.

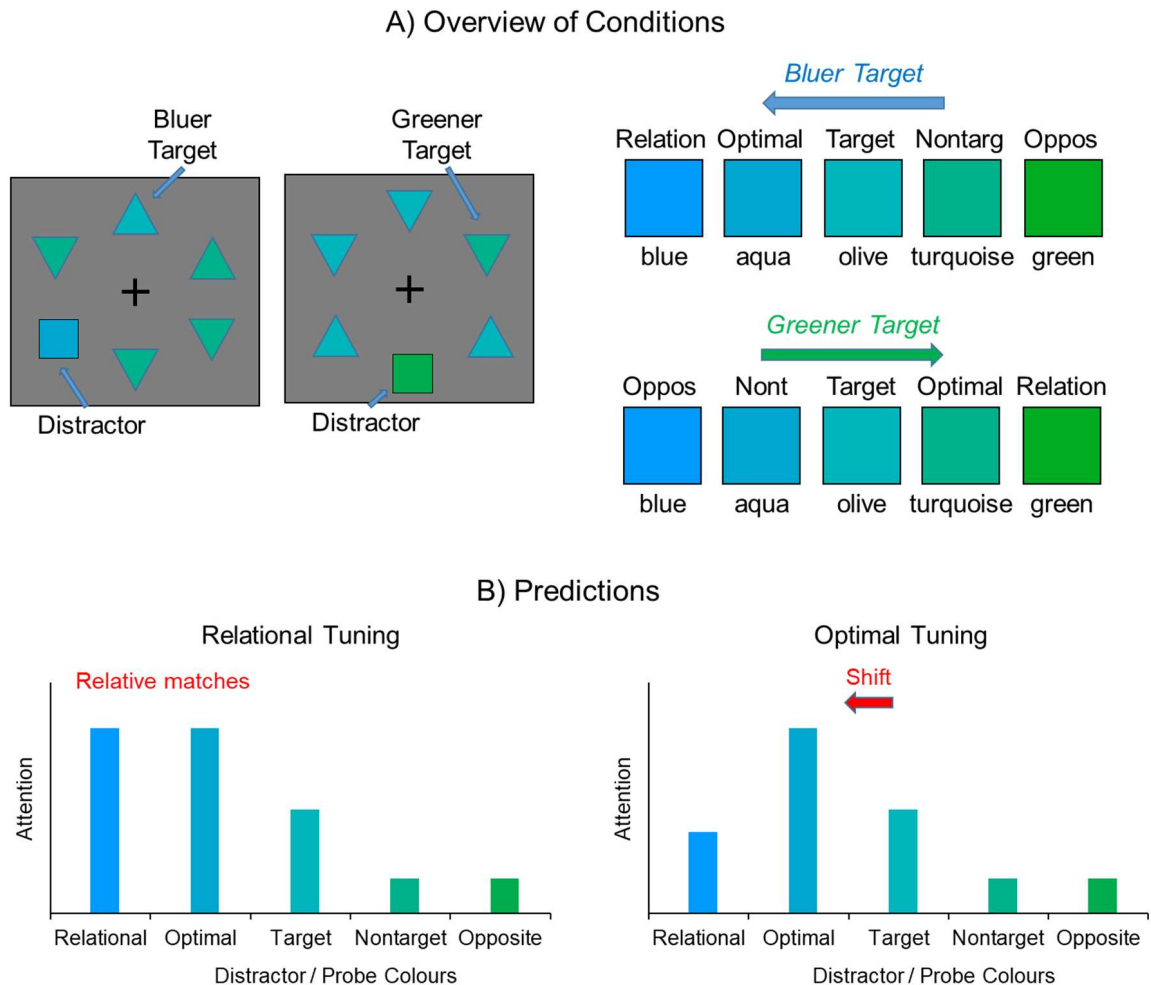

*Figure 1.* A) The top left shows example search displays for each condition (bluer, greener target). Participants had to report the direction of the olive target triangle while ignoring the square distractor. The left panel shows an example of the midline target/laterlised distractor condition, whereas the right panel shows an example of the midline distractor/lateralised target condition. The top right shows the colours used as distractor and probe colours and the associated labels when the olive target was presented among turquoise nontargets (bluer target condition) vs. aqua nontargets (greener target condition). B) Predictions of the relational tuning account and optimal tuning account: According to the relational account, attention should be tuned to the relative colour of the target (e.g., bluer) and as a consequence, all distractors that are bluer than the target should attract attention most strongly (left). According to optimal tuning, attention should be tuned to a feature value that is slightly shifted away from the nontargets when the target Is "Imil'r to the nontargets. In this case, a target-coloured 'exaggerated' target colour should attract attention most strongly, whereas relatively matching dissimilar colours should not attract attention.

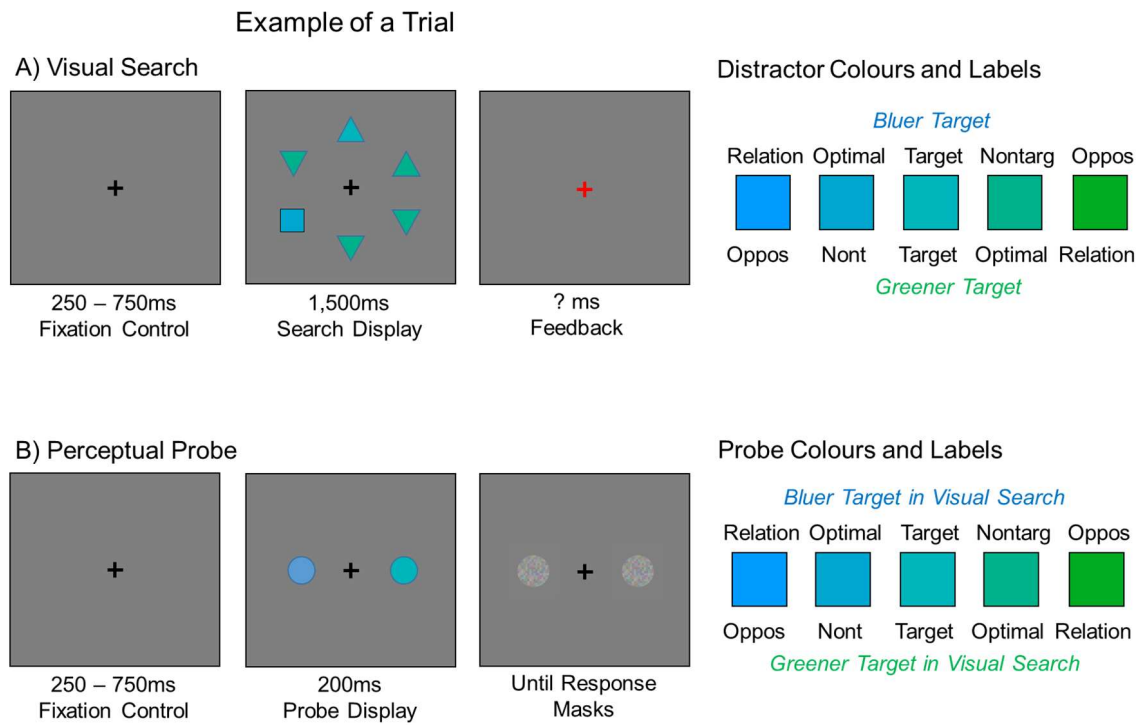

*Figure 2.* Overview of the scheduling of events and example displays in A) the visual search task and B) the perceptual probe task.

If covert attention behaves similarly to overt attention (i.e., eye movements) in the adapted displays, we would expect optimal and relatively matching distractors to attract attention equally strongly, reflected in equally large N2pc amplitudes to optimal and relational distractors (which should be larger than the N2pc amplitude to the target-coloured distractor; see Fig 1B, left). In turn, the accuracies in the probe task should show a significantly larger impairment in the presence of an optimal distractor than a relational distractor (see Hamblin-Frohman & Becker, 2021; Yu et al., 2022).

**Method**

*Participants.* To determine the required sample size for the study, we used the difference in N2pc amplitude between a target-matching and a relatively matching distractor, as observed in Martin & Becker's (2018) EEG study ($\eta_p^2 = .32$). The BUCSS R package power analysis tool (Anderson et al., 2017). The analysis suggested a sample size of 37 participants for 95% power (with 75% assurance).

Forty-one participants were recruited with self-reported normal or corrected-to-normal vision. Participants were compensated with either AU$40 or course credit for participating in the experiment. Three participants were excluded from further analyses

as more than 50% of trials were lost due to channel noise or excessive eye movements in the EEG. This left 38 participants in the final analysis ($M$ age = 21.3, SD = 4.33, 27 female).

*Apparatus.* The current study used a 32-channel BrainProducts EEG system (Gilching, Germany, 2016) with a BrainAmp DC amplifier connected to a personal computer (Intel Core i5-4790 3.50 GHz processor) with an Intel ®HD Graphics 4600 graphics card. Participants' eye movements were monitored through the SR-Research EyeLink 1000 remote eye tracker at a 500 Hz sampling rate. The experiment was controlled by PsychoPy2 (Peirce, 2007), connected to a 19" colour LCD monitor with a resolution of 1,280 x 1,024 pixels and a 60 Hz refresh rate. The viewing distance between the participants and the display monitor was approximately 55 cm. All participants individually participated in the experiment in a dimly-lit EEG laboratory.

*Stimuli.* The stimuli in the visual search task were presented against a grey background (RGB: [127.5, 127.5, 127.5]). Each trial began with a black fixation cross (height: 0.31°) presented in the centre. The search display consisted of six triangles (width: 1.88°, height: 1.98°) or five triangles and one distractor square (1.77° x 1.77°) at 4.79° from fixation. The target was always olive, and participants were randomly assigned to search for the olive target among either bluer non-targets (completing a *greener* target search),  or greener non-targets (completing a *bluer* target search; see Figure 2). The distractor square could have five possible colours ([blue: 17, 169, 175], [aqua: 37, 170, 161], [olive: 56, 171, 146], [turquoise: 74, 171, 131], [green: 90, 170, 116]), which were re-coded as relational, optimal, target(-coloured), nontarget(-coloured) and opposite distractor, depending on the search condition (see Fig. 1A & 2).

The stimuli in the perceptual probe task (*probes*) consisted of two differently coloured circles (radius: 1.25°), presented 4.37° to the left and right from fixation, against the same grey background as used in the visual search task. The probe colours always consisted of one target colour (probe target) and one of the four possible distractor colours from the visual search task (see Fig.2 ; relationally matching, optimally coloured, nontarget colour, and relationally opposite colour). The probes would only appear briefly and were then backward-masked with same sized circles with a colour checkerboard pattern.

*Design.* The experiment employed a mixed design, with target condition (greener, bluer) varying between participants, and distractor / probe conditions varying within participants. There were 640 lateralised distractor trials (128 per distractor colour) in visual search, evenly distributed between distractor presented on the left or right of fixation. To prevent that the target position became predictable, we included 64 foil trials, where the distractor was presented on the midline with a lateralised target, and 512 trials, in which the target was lateralised and there was no distractor. These trials were not analysed, as we were primarily interested in assessing the N2pc to the different distractors (which requires a lateralised distractor). The orientation of the target and non-target triangles varied randomly in each display.

We also included 512 perceptual probe trials, in which the target coloured probe was presented along with one of the four distractor colours (128 trials per distractor colour). On half of the trials, the target coloured probe was presented on the right side of the screen, and on the other half, on the left. Probe trials were randomly interleaved with visual search trials. Before commencing the main experimental block, participants completed 30 no-distractor visual search trials, then four distractor-present trials and two probe trials as practice (not analysed).

*Procedure.* For the visual search trials, participants were instructed to search for the uniquely coloured triangle (olive) and report its orientation with the left arrow key (triangle-down) or right arrow key (triangle-up) as quickly as possible. Moroever, participants were instructed that they should try to ignore the distractor squares, as they were irrelevant. For the perceptual probe trials, participants were instructed to report the location of the target-coloured probe by pressing either the left or right arrow key on the keyboard, dependant on its location.

Prior to the experiment, participants completed a 9-point calibration. Gaze was continually monitored throughout each trial. Participants were instructed to keep their gaze on the central fixation cross throughout the entire experiment. If gaze left this region (radius: 1.75°), a feedback message was displayed at the end of the trial reminding participants to maintain fixation on the fixation cross. After five gaze violations, the eye-tracking calibration procedure was re-run.

Each display began with a fixation cross presented for a variable duration between 750ms and 1250ms. The search items were displayed for 1500ms, or until a response was made. If a response was not made in this period a "too slow" message was displayed as feedback. If an incorrect response was made, the fixation cross flashed red for 200ms. Both types of error trial were excluded from analysis.

On perceptual probe trials, the two coloured probes were presented after the fixation period (750 – 1250m) for 250ms and were then backward-masked by the checkerboard pattern masks. The masks remained on the screen until a response was recorded. There were no time limits for the probe trials, and there was never any accuracy feedback.

*EEG Data Recording and Analysis.* The continuous EEG data was recorded from 29 scalp electrodes mounted in an elastic cap (Fpz, F7, F3, Fz, F4, F8, FC5, FC6, T7, C3, Cz, C4, T8, CP5, CP6, P7, P3, Pz, P4, P8, PO9, PO7, PO3, PO4, PO8, PO10, O1, Oz, and O2). The impedance level was kept below 10 kΩ. All electrodes were referenced online to the left earlobe and offline to the average of all electrodes. The data sample rate was 500 Hz, and the online high cut-off filter rate was 40 Hz, paired with a 50 Hz notch filter. In order to prevent eye movements and eye blinks from contaminating the EEG data, all trials where the Horizontal Electrooculogram (HEOG) and muscle artefacts amplitudes exceeded ±80 μV were excluded. The remaining data were segmented into epochs starting from 100 ms before stimulus onset to 600 ms post-stimulus onset, and baseline-corrected using the 100 ms pre-stimulus interval.

EEG data were analysed using EEGLAB (Delorme & Makeig, 2004) and ERPLAB (Lopez-Calderon & Luck, 2014). To identify the electrode pair that would produce the most diagnostic signals for the assessment of the N2pc, we computed the mean N2pc amplitude (contra minus ipsilateral waveforms) for the lateralised target condition (with mid-line distractor) separately for all available posterior electrodes (P3/4, P7/8, PO3/4, PO7/8, PO9/10, and O1/2), 200-300 post-stimulus. The results showed a significant N2pc to the target in all electrode pairs, all $ts>6.1$, all $ps<.001$, with the largest mean difference detected at electrodes PO3/4 (mean difference: 0.97 μV). Hence, the analysis of the effects of the lateralised distractors was based on electrodes PO3/4 (which were also used in previous research; see e.g., Oemisch et al., 2017; Foster et al., 2020; Papaioannou & Luck, 2020.

**Results**

*Data.* In the visual search task, the accuracies were very high and quite similar across all conditions (M: 93.9%; range: 92.5% - 94.7% on average in each condition), and hence, the analysis of visual search performance focussed on RT rather than accuracies. In the perceptual probe task, we only analysed the mean accuracies and not the mean RT, as participants were instructed to respond accurately without any time restrictions (see Hamblin-Frohman & Becker, 2021, Navalpakkam & Itti, 2007; Yu et al., 2022, for the same reporting procedures).

The behavioural data and EEG results were analysed using repeated-measures analyses of variance (ANOVAs) and two-tailed, pairwise *t*-tests. For the ANOVAs, the Greenhouse-Geisser corrected *p*-values and effect sizes were reported where appropriate, together with the uncorrected degrees of freedom (*df*s). All data were analysed using SPSS 29 statistical software (IBM).

### Visual Search RT.

*Mean RT.* The mean RT of participants' correct responses were first analysed with a 2 x 6 ANOVA with the between-subjects factor 'target condition' (bluer, greener target) and the within-subjects factor of 'distractor' (relational, optimal, target, nontarget, opposite, absent). The results revealed a significant main effect of distractor condition, $F(5, 180)=91.5$, $p<.001$, $\eta_p^2=.72$ (Greenhouse-Geisser corrected), but no significant main effect of the target condition and no interaction between the two variables, both *F*s<1. Hence, for the subsequent analyses, data were pooled over the greener and bluer target conditions (while we still displayed the data separately in Figures 3 and 4).

Two-tailed, pairwise *t*-tests revealed significantly longer RTs in the relational distractor condition than the optimal distractor condition, $t(37)=2.9$, $p=.003$, $\eta_p^2=.18$, and longer RT in the optimal than the target-coloured distractor condition, $t(37)=7.3$, $p<.001$, $\eta_p^2=.59$. Thus, in line with the relational account, both the relational and optimal distractors delayed responses more than the target-coloured distractor, indicating that they both strongly attracted attention. Contrary to the optimal tuning account, the relational distractor was not less effective than the optimal distractor, but interfered slightly more than the optimal distractor.

The target-coloured distractor also delayed responses more than the nontarget-coloured distractor, $t(37)=9.4$, $p<.001$, $\eta_p^2=.70$, as predicted by all theories, and responses were slightly faster with the nontarget-coloured distractor than with the opposite distractor, $t(37)=2.7$, $p=.012$, $\eta_p^2=.16$, which in turn showed shorter RTs than the distractor absent control condition, $t(37)=4.0$, $p<.001$, $\eta_p^2=.31$.

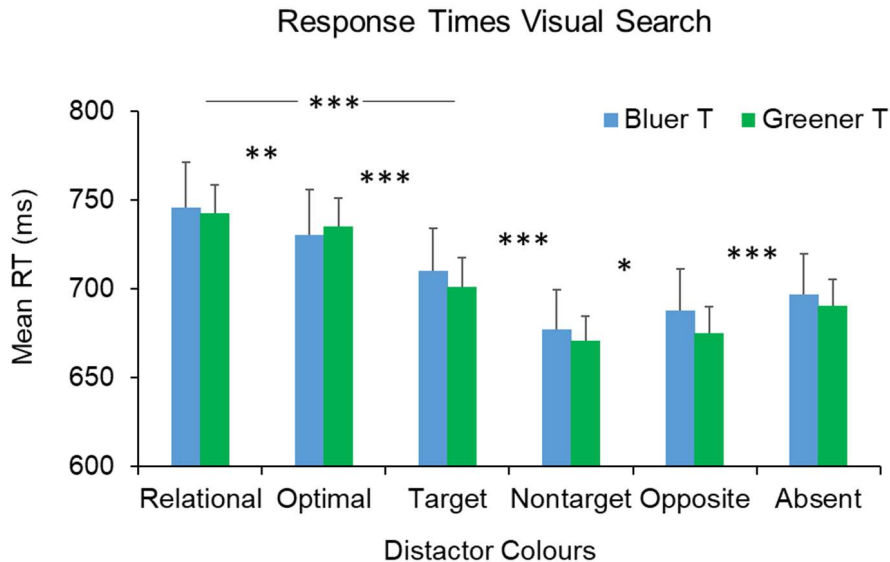

*Figure 3.* Mean RTs in visual search, depicted separately for the six different distractor conditions and for the different target conditions (bluer, greener target). Results did not differ between the two target conditions. As predicted by the relational account, both the relational and optimal distractors delayed responses more than the target-coloured distractor, indicating that they both strongly attracted attention. The target-coloured distractor also slowed RTs more than the nontarget-coloured distractor. The nontarget-coloured distractor in turn slightly speeded RTs compared to the opposite distractor, which led to faster RTs than observed in the distractor absent control condition. Error bars represent the mean Standard Error of the Mean (SEM). *$p$<.05, **$p$<.01, ***$p$<.001 (as per two-tailed *t*-test).

### Probe Task Errors

As participants were instructed to respond accurately, only the mean error scores were analysed to assess participants' perceptual judgement for the target colours. A 2 x 4 ANOVA with the between-subjects factor target condition (bluer, greener target in visual search) and the within-subjects variable probe colour (relational, optimal, nontarget-coloured, opposite) was computed over the mean errors in the probe task. The results showed a significant main effect of probe colour, $F(3,108)=35.7$, $p$<.001, $\eta_p^2=.50$, whereas target condition had no effect and did not interact with the probe colours, $F$s<1.2, $p$s>.31. Hence, data were pooled over target conditions for the subsequent analyses.

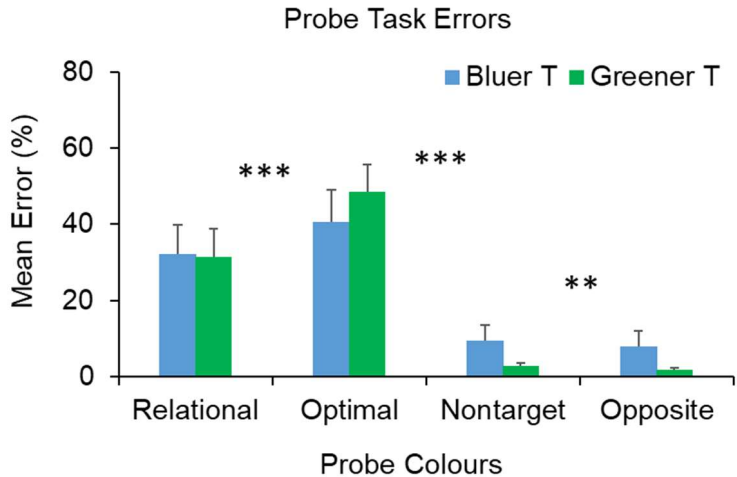

*Figure 4.* Mean errors in the perceptual probe task, depicted separately for the four probe colours that were presented along with the target-coloured probe (relational, optimal, nontarget and opposite coloured probes). The optimally-coloured probe led to the most frequent errors, significantly more errors than the relational or nontarget-coloured probes, in line with the optimal tuning account. Participants also committed slightly more errors with the nontarget-coloured probe than with the opposite coloured probe. Error bars represent the Standard Error of the Mean (SEM). **$p$<.01, ***$p$<.001, as per two-tailed *t*-test.

As shown in Figure 4, pairing the target colour with the optimal colour led to the highest error scores. In line with the optimal tuning account, significantly more errors were committed with the optimal probe than with the relational probe, *t*(37)=6.4, $p$<.001, $\eta_p^2$=.50, or the nontarget-coloured probe, *t*(37)=7.1, $p$<.001, $\eta_p^2$=.57. In addition, the presence of the nontarget-coloured probe evoked slightly less errors than the opposite coloured probe, *t*(37)=2.8, $p$=.004 $\eta_p^2$=.18.

**EEG: N2pc**

Figure 5 shows the grand mean difference waveforms (contralateral minus ipsilateral) for each of the distractors, -100 to +400ms from stimulus onset. As shown in the graph, the waveforms for the target-coloured distractor seemed to follow a slightly different time-course than for the relational distractor. We addressed this potential problem by first analysing the N2pc and Pd in fixed time windows that were identical to those used in a previous, similar study (Martin & Becker, 2018), and then providing a more fine-grained analysis of the results, where we averaged the waveforms across consecutive 20ms time bins, and compared the binned results across the different distractors (see Fig. 6).

For the analysis of the N2pc, we first conducted a 2 (between-subjects target colour condition: bluer, greener) x 5 (within-subjects distractor condition: relational, optimal, target, nontarget, opposite) mixed ANOVA over the mean amplitudes of the difference waves (contralateral minus ipsilateral) in the time window of 220 – 280 ms post-stimulus (the same time window as used in Martin & Becker, 2018). The results revealed a significant main effect of distractor condition, $F(4, 144) = 15.3$, $p<.001$, $\eta_p^2 = .30$, but no main effect of target condition or an interaction, both $F$s<1. Hence, for subsequent analyses, the data were pooled over the target colour conditions.

To assess which distractors produced a significant N2pc, we compared the mean amplitudes of the difference waveforms (contralateral minus ipsilateral) against zero. A significant N2pc was found for the relational distractor, $t(37)=3.9$, $p<.001$, $\eta_p^2=.29$, the optimal distractor, $t(37)=4.2$, $p<.001$, $\eta_p^2=.33$, and the target-coloured distractor, $t(37)=3.0$, $p=.003$, $\eta_p^2=.19$. By contrast, the nontarget-coloured and opposite distractors did not show a significant N2pc, but a contralateral positivity (i.e., Pd), which was significant for the opposite distractor, $t(37)=3.2$, $p=.002$, $\eta_p^2=.21$, but not for the nontarget-coloured distractor, $t<1$.

Pairwise, two-tailed comparisons revealed that the N2pc of the relational and optimal distractor did not differ, $t<0$. However, the optimal distractor had a significantly larger N2pc than the target-coloured distractor, $t(37)=2.4$, $p=.019$, $\eta_p^2=.14$, which in turn had a significantly larger N2pc than the nontarget-coloured distractor, $t(37)=3.2$, $p=.003$, $\eta_p^2=.21$. The nontarget-coloured distractor and opposite distractor did not differ significantly from each other, $t(37)=1.7$, $p=.090$.

The results of this N2pc analysis support the relational account, as both the relational and optimal distractors were selected equally often, and attracted attention more strongly than the target-similar distractor. However, this result still needs to be confirmed by a more fine-grained analysis (see below).

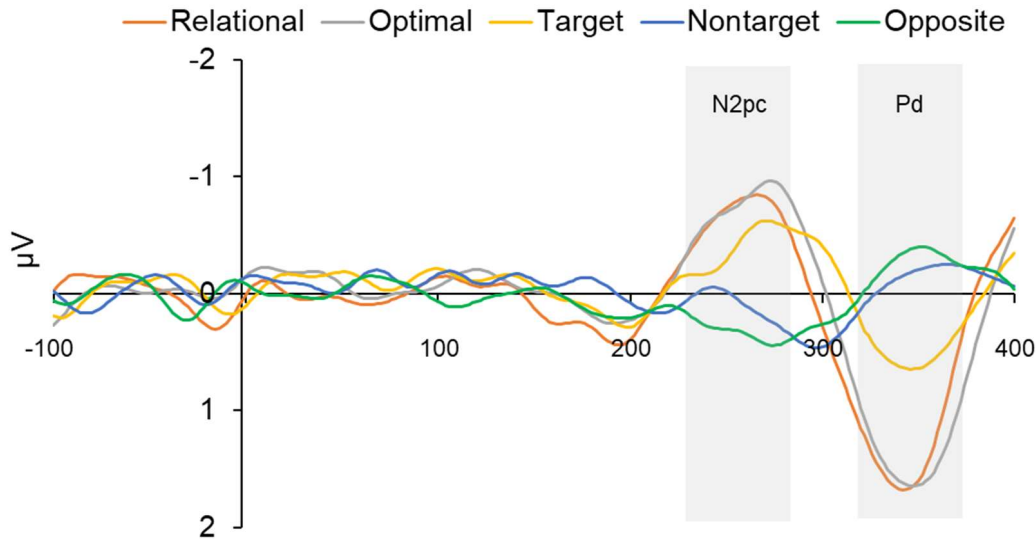

*Figure 5.* Difference waveforms (contra-minus ipsilateral waveforms) from electrodes PO3/4, depicted separately for the different distractor conditions (relational, optimal, target-coloured, nontarget-coloured, and opposite). The results show a large N2pc for relational and optimal distractors, and a slightly but significantly smaller N2pc for target-coloured distractors in an early time window (220-280 ms). This is followed by a pronounced Pd to relational and optimal distractors, which was significantly larger than the Pd for the target-coloured distractor (which also did not differ significantly from zero). The nontarget-coloured and opposite distractors did not elicit a significant N2pc.

**EEG: Pd**

To analyse the data for a possible late Pd, we first computed a 2(target colour) x 5(distractor colour) mixed ANOVA over the mean amplitudes of the difference waves (contralateral minus ipsilateral) in the time window 310 – 370 ms post-stimulus at electrodes PO3/4. The results revealed a significant main effect of distractor condition, $F(4,144)=12.3$, $p<.001$, $\eta_p^2=.25$, but not of target condition, $F(1,35)=2.2$, $p=.15$, and the interaction was also not significant, $F<1$. Thus, we pooled the data over the two target conditions for the following analyses.

Two-tailed *t*-tests revealed a significant Pd only for the relational and optimal distractors, $t(37)=7.9$, $p<.001$, $\eta_p^2=.63$, and $t(37)=5.5$, $p<.001$, $\eta_p^2=.45$; not for any of the other distractors (target-coloured, nontarget-coloured or opposite), all *t*s<1.2, *p*s>.23.

Comparing the magnitude of the Pd across the different distractors showed equally large Pds for the relational and optimal distractors, *t*<0. However, the optimal distractor generated a significantly larger Pd than the target-coloured distractor, $t(37)=3.9$, $p<.001$, $\eta_p^2=.29$, which in turn did not differ significantly from the nontarget-coloured distractor, $t<1$. The nontarget-coloured distractor also did not differ from the opposite distractor, $t<1$.

### EEG: N2pc bins (200 – 300 ms)

Figure 5 indicates that the relational, optimal and target-similar distractor have slightly different time courses, with the N2pc of the target-coloured distractor showing some delays, compared to the relational and optimal distractors. To analyse possible differences in the time-course of the distractors in more detail, we averaged the difference waveforms (contralateral minus ipsilateral) across consecutive time bins and compared the effects of the different distractors within each bin. For better readability of the results, we included only significant differences in the description of results, and reported only the $p$-values of the two-tailed $t$-tests (data available upon request).

In the first bin (200 – 220ms), none of the distractors showed a significant N2pc or Pd, as none of them differed significantly from zero, all $ps>.17$. In the second bin (220 - 240ms), a significant N2pc began to emerge for the relational distractor, $p=.021$, and the optimal distractor $p=.041$, whereas the other distractors did not differ significantly from zero, $ps>.24$. In the third bin (240 – 260ms), the relational, optimal and target-similar distractor all showed a significant N2pc, $ps\leq.01$, and the N2pc was significantly larger for the relational and optimal distractors than for the target-similar distractor, both $ps<.041$. A significant Pd started to emerge for the opposite distractor, $p=.007$. In the fourth bin (260 – 280 ms), the N2pcs for relational, optimal and target-similar distractor were at a peak, all $ps<.001$, with the optimal distractor showing the largest N2pc amplitude, which was significantly larger than the N2pc of the target-coloured distractor, $p=.030$. Both the nontarget-simlar and opposite distractor showed a significant Pd, $ps<.001$. In the last time bin for the N2pc (280 – 300ms), the N2pc for the relational distractor had already declined to be indistinguishable from zero, $p=.18$, while the optimal and target-similar distractor still showed a significant N2pc, $ps\leq.01$. The nontarget-simlar and opposite distractor were both still showing a solid Pd, $ps<.003$. These results support a relational account, as the relational and optimal distractors did not differ, and were equally large or larger than the N2pc of the target-coloured distractor.

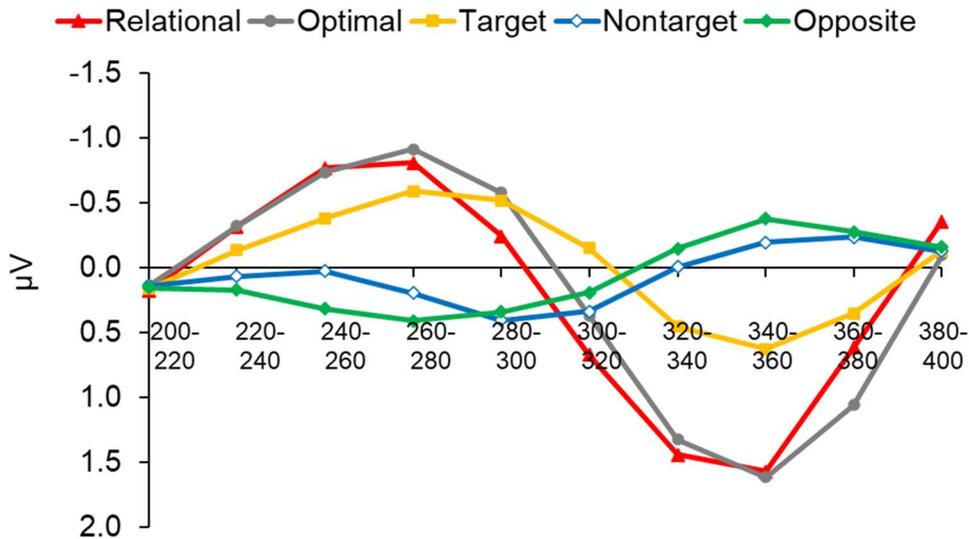

*Figure 6.* Difference waveforms (contra-minus ipsilateral) from averaged electrodes PO3/4, depicted separately for the different distractor conditions (relational, optimal, target-coloured, nontarget-coloured, and opposite

**EEG: Pd bins (300 – 400 ms)**

In the first bin of the late Pd time window (300 – 320 ms), both the relational and optimal distractor show a significant Pd, $p<.001$ and $p<.024$, respectively, while the target-similar distractor still produced a negative deflection, which did not differ from zero, $p=.36$ (but differed significantly from both relational and optimal distractors, $p$s≤.015). The waveform for the nontarget-similar distractor continued to show a significant Pd, $p=.002$, while the Pd of the opposite distractor did not differ from zero, $p=.14$. In the second bin (320 – 340 ms), the relational, optimal and target-coloured distractor all showed a significant Pd, all $p$s≤.001, whereby the Pd of the relational and optimal distractor were significantly larger than the Pd for the target-similar distractor, both $p$s<.001. The waveforms for the nontarget-similar and opposite distractor were now slightly negative and indistinguishable from zero, $p$s>.24. In the third bin (340 – 360 ms), the Pds for the relational, optimal and target-similar distractor were at a peak, all $p$s<.001, and the Pds for the relational and opposite distractors were significantly larger than for the target-similar distractor, $p$s<.001. The opposite distractor showed a significant negative deflection, $p=.005$. In the fourth bin (360 – 380ms), the Pds for the relational, optimal and target-similar distractor were still significant (all $p$s≤.01). However, the Pd for the relational distractor was already strongly reduced and significantly smaller than the Pd for the optimal distractor, *p*=.026. The Pd of the optimal distractor was now larger than the Pd for the relational and target-similar distractor, *p*=.026 and *p*≤.001, respectively. In the last bin of the late Pd (380 – 400ms), all waveforms returned to zero or slightly negative values, wherbey only the relational distractor differed significantly from zero, *p*<.044. In sum, the more fine-grained analysis revealed a significant Pd also for the target-coloured distractor, 320 – 380ms post-stimulus, whereby this Pd was however smaller than the Pds of the relational and optimal distractors.

## General Discussion

The results of the present study largely replicated previous results of eye movement studies (Hamblin-Frohman & Becker, 2021, Yu et al., 2022), and confirm that these previous findings can be extended to covert attention. Hamblin-Frohman and Becker (2021) were the first to report that early selection operates on relative features, as evidenced by the results of the first eye movement on a trial, whereas performance in perceptual judgement tasks follows the optimal tuning account, as reflected in probe task results. They proposed that optimal tuning does not describe early attentional selection, but later perceptual decision-making processes. Both the findings and general conclusions were later confirmed by Yu et al. (2022), who conducted a similar study using eye movements to index attentional guidance.

The results of the present study show that the same conclusions can be drawn in tasks not allowing any eye movements. Using the N2pc in the EEG of participants to assess covert attention revealed that attention was equally strongly attracted to relationally and optimally coloured distractors, in line with the relational account. However, the perceptual probe task results revealed significantly more errors with the optimal distractor than the relational distractor, reflecting that only the optimal distractor was likely to be confused with the target. This means that early attentional selection is biased towards all items that match the relative feature of the target, while perceptual decision-making is biased more narrowly towards a slightly shifted, 'exaggerated' target feature value, as predicted by optimal tuning. This dissociation between the visual search results and perceptual probe results suggests that early selection and later, perceptual decision-making are based on different target templates or differences in how attention is tuned to the target (cf. Hamblin-Frohman & Becker, 2021; see also Yu et al., in press).

While the mean N2pc amplitudes of the relational and optimal distractor did not differ, the mean RTs in visual search showed slightly longer RTs in the presence of a relational distractor than optimal distractor. This result pattern has often been observed in previous studies on the relational account, including in eye movement results (e.g., Becker, 2010). It is likely that this is due to the greater distinctiveness of the relationally matching distractor: As this distractor is more dissimilar from the target, it is more likely to register as 'the bluest item in the visual field' than the optimal distractor, which is more similar to the target (and even confusable with the target; see Fig. 4). Another explanation is that the relationally matching distractor is more salient than the optimal distractor, as it has a slightly higher feature contrast to the other items (target and nontargets) than the optimal distractor. Previous studies have included a very salient and highly dissimilar distractor as a control and did not find any strong effects for this distractor (e.g., Becker, Lewis & Axtens, 2017; Martin & Becker, 2018; York & Becker, 2020), arguing against a saliency explanation. However, as the two explanations have never been formally tested, they both remain possible (and other explanations are conceivable as well).

The present findings may also inform the current debate about the N2pc. It has been argued that the N2pc does not reflect covert attention shifts, but later, attentional engagement or stimulus processing after attention has been allocated to a stimulus (e.g., Ramgir & Lamy, 2022; Zivony et al., 2018). Previous studies showed that early visual selection (as indexed by the first eye movement) was driven by relational tuning, while later, perceptual decision-marking followed optimal tuning (Hamblin-Frohman & Becker, 2021; Yu et al., 2022). Hence, if the N2pc reflected a late component of attentional engagement or feature analysis concerned with decision-making (e.g., whether a selected item is the target or not), we would have expected the N2pc results to be more aligned with optimal tuning, not relational tuning. The fact that the N2pc results were more closely aligned with prior eye movement results thus shows that it reflects an early component of attentional engagement rather than perceptual decision-making about whether a selected item is the target or not.

The present study also allows some interesting new insights into the Pd. In previous studies, the Pd has been reported to occur either in an early time window (similar time window as the N2pc), or in a later time window, often following a significant N2pc to a distractor (e.g., Kiss et al., 2012; Sawaki & Luck, 2010; Sawaki, Geng & Luck, 2012). Here, we observed both an early and late Pd. The opposite distractor, which had an extreme, relatively non-matching colour (e.g., green in search for a bluer target), showed a significant Pd in the early time window, in which the relational, optimal and target-coloured distractor showed a significant N2pc. This could indicate that tuning attention to the relative feature of the target led to automatic suppression of items that differ in the opposite direction from the other search items, prior to selection.

In turn, the relational and optimal distractors elicited a significant Pd after selection, which followed a significant N2pc in the earlier time window (see Fig. 5). This Pd most likely reflects inhibition of the relational and optimal distractors after selection, to continue search for the target. The target-coloured distractor also showed a positive deflection in the Pd time-window which was however significantly weaker than the Pd of the relational and optimal distractors. The target-coloured distractor may have showed weaker inhibition because it had the same colour as the target, limiting the ability to inhibit the distractor. If this is the case, inhibition of the distractor may not be mediated by the simple act of detecting that it is a distractor, but (also) by the colour of the distractor (i.e., mediated by feature-based processes). This is in line with previous accounts of feature-based inhibition (e.g., Treisman & Sato, 1990) and current accounts, such as the revised *signal suppression hypothesis* (e.g., Gaspelin & Luck, 2019).[1]

It should be noted, though, that the interpretation of the Pd as reflecting inhibition is still contentious. It is equally possible that the Pd merely reflects an imbalance in the distribution of attention across the two visual fields. In the case of the late Pd that followed an N2pc, the Pd may reflect that attention is more likely to be shifted to the other side of the display in search for the target, after the distractor was rejected (e.g., Kerzel, & Burra, 2020). Similarly, the early Pd observed for the opposite distractor may indicate that attention was more likely to be shifted to the side opposite of the distractor, as the distractor was relationally less matching than the nontargets, providing a competitive advantage to the stimuli on the other side of the display (e.g., Schoenhammer et al., 2020). While this question remains to be addressed in future research, the results of the present study clearly showed an early Pd in the N2pc time window for a distractor with a relationally opposite colour to the target, and a late Pd following the N2pc in relatively matching distractors, demonstrating that both early and late Pds can be observed in the same data set (for different distractors).

---

[1] While the signal suppression hypothesis was originally formulated to suggest suppression of all saliency signals, regardless of their origin (e.g., Sawaki & Luck, 2010), it was later revised to suggest that suppression of a salient item was mediated by suppressing its feature (e.g., Gaspelin & Luck, 2019).

In this respect, it is perhaps interesting to note that the Pd was quite large in the present study, peaking at approximately 1.5 µV. This seems larger than the Pd reported in previous studies, which peaked at approximately 0.5 µV (e.g., Drisdelle & Eimer, 2021; Gaspar & McDonald, 2014; Sawaki & Luck, 2010). It is possible that, in order to observe a large Pd, it is necessary to use relationally opposite stimuli (for an early Pd) or relatively matching stimuli that very strongly attract attention (for a late Pd).

Last but not least, is also interesting to note that the time-course of the relational, optimal and target-coloured distractors showed some differences (see Fig. 6). The N2pc for the relational distractor seemed to peak slightly earlier and disintegrate earlier than the N2pc for the optimal distractor, while the target-coloured distractor seemed to have the longest-lasting effects within the N2pc time window.

At first, these results may be taken to show that the N2pc may also reflect the speed of distractor rejection, and thus, decision-making, in the decline or offset of the N2pc (after the peak). However, in the present data set, the N2pc to the relational and optimal distractors was swiftly followed by a large Pd, so that it is possible and perhaps even more likely that the speed of distractor rejection is reflected in the time-course of the Pd. Of note, the relational distractor seemed to show a slightly earlier Pd than the optimal distractor or target-coloured distractor, which again both had longer-lasting effects. This results pattern is reminiscent of previous eye tracking studies, which showed longer dwell times for target-similar distractors than for the target-dissimilar, relational distractor (e.g., Becker et al., 2014; Martin & Becker, 2018), presumably reflecting that target-dissimilar distractors can be more rapidly rejected after selection than target-similar distractors, which require a more in-depth feature analysis (e.g., Becker, 2011). This indicates that the N2pc and Pd may also be sensitive to target similarity – a factor that does not seem to play a role in the initial allocation of attention, but one that has large implications for the speed of distractor rejection and other decision-making processes. If this can be corroborated in future studies, it would mean that the N2pc/Pd would not reflect solely the initial allocation of attention to a stimulus, or the initial attentional engagement, but could also index the time-course of distractor rejection and disengagement.

To summarise, the results of the present study show that covert attention behaves similarly regardless of whether a task requires eye movements or allows only covert attentional selection, and covert attention is indexed by the N2pc. Despite differences in the stimulus displays, we found that covert attention was allocated to all relatively matching distractors, in line with the relational account, whereas later perceptual decision-making processes followed the optimal tuning framework. These results reinforce the view that the N2pc mainly indexes processes of attentional selection or early attentional engagement (which depend on relative matches) and argue against the notion that it predominantly indexes later processes concerned with decision-making (which depend on a feature match with the target). The results also showed a large late Pd after selection of relatively matching distractors, and an early Pd in response to a relationally opposite distractor, suggesting that the Pd to a stimulus depends on feature-based attention and may be modulated by target similarity.

**Author Note.**

This research was supported by the Australian Research Council Discovery Grant DP DP210103430 to SIB and UQ PhD scholarships to ZQ.